\documentclass[11pt]{article}

\usepackage[margin=0.85in]{geometry}
\usepackage{booktabs}
\usepackage{longtable}
\usepackage{array}
\usepackage{ragged2e}
\usepackage[breaklinks=true,colorlinks=true,linkcolor=black,urlcolor=blue,citecolor=black]{hyperref}

\usepackage{xcolor}
\usepackage{enumitem}
\usepackage{amsmath}
\usepackage{caption}
\usepackage{titlesec}
\usepackage{parskip}
\usepackage{fancyhdr}
\usepackage{graphicx}

\newcolumntype{L}[1]{>{\RaggedRight\arraybackslash}p{#1}}

\title{\textbf{Choosing a Text Embedding Model: A Practical Benchmarking and Decision Framework}\\[0.3em]
\large Retrieval Performance, the MTEB Landscape, and End-to-End Deployment Considerations}

\author{Madhav S Baidya \\[2pt]
\normalsize Indian Institute of Technology (BHU) Varanasi \\[1pt]
\normalsize \texttt{madhavsukla.baidya.chy22@itbhu.ac.in}}

\date{\today}

\begin{document}

\maketitle

\begin{abstract}
\noindent This report develops a practical, evidence-based framework for embedding model selection, built on a benchmarking study that evaluates T3EM (Text 3 Embedding Model), a commercial API-based embedding model, against a broad set of open-source alternatives on English-language retrieval tasks, and situates these findings within the wider Massive Text Embedding Benchmark (MTEB) landscape spanning classification, clustering, semantic similarity, reranking, pair classification, bitext mining, and summarization. Beyond raw benchmark scores, the report traces the full path from embedding model to retrieved result --- how embeddings are produced, how they are indexed and searched at scale, and how document chunking strategy shapes retrieval quality --- so that model choice can be reasoned about as one decision within a complete retrieval pipeline rather than in isolation. The result is a consolidated set of practical recommendations for selecting an embedding model according to task, latency, cost, and deployment constraints.
\end{abstract}

\vspace{0.5em}
\begin{center}
\fcolorbox{black}{gray!10}{%
\begin{minipage}{0.92\textwidth}
\textbf{Executive Summary}
\begin{itemize}[nosep]
    \item T3EM (Text 3 Embedding Model) achieves the highest retrieval quality of any model evaluated (average nDCG@10 = 0.638), but at roughly 7--14$\times$ the latency of open-source alternatives and a per-query API cost they do not carry.
    \item mE5-L is the strongest open-source alternative for general-purpose English retrieval, scoring within 0.09 nDCG points of T3EM at a fraction of the latency and no per-query cost --- and is the recommended default when requirements are unspecified.
    \item Models trained for sentence similarity rather than retrieval (LaBSE, mMPNet) substantially underperform on retrieval tasks despite scoring competitively on similarity benchmarks: training objective, not model size, is the dominant factor in retrieval quality.
    \item No single model wins across all task types: ST5 leads on semantic similarity, MPNet leads on clustering and reranking, GTR and SGPT variants lead on broader MTEB retrieval, and LaBSE leads on cross-lingual bitext mining.
    \item Chunking strategy matters: quality plateaus by a chunk size of 32 tokens, semantic chunking meaningfully outperforms fixed-size chunking at small chunk sizes, and quality collapses for every model below roughly 16 tokens per chunk.
\end{itemize}
\end{minipage}%
}
\end{center}
\vspace{0.5em}

\vspace{0.3em}
\vspace{0.5em}

\tableofcontents
\newpage

\section{Introduction}

\subsection{Motivation and Objective}

This study was undertaken to evaluate the retrieval performance of T3EM (Text 3 Embedding Model), a commercial API-based embedding model, against a range of widely used open-source embedding models on English-language retrieval tasks. The effect of document chunking strategy on retrieval quality was also examined. The purpose of the work was to determine under which conditions the additional cost and latency of a commercial API is justified, and under which conditions an open-source alternative performs comparably.

To situate these retrieval-specific findings within a broader picture of embedding model behavior, this study also draws on the Massive Text Embedding Benchmark (MTEB), which evaluates a much larger pool of models across eight distinct downstream task categories. Combining the two sources makes it possible to compare directly measured retrieval results against a wider landscape of published, task-diverse performance data, using a single consistent set of datasets, models, and metrics rather than treating them as separate studies.

\subsection{What Is a Text Embedding?}
\label{sec:embedding-definition}

A text embedding model converts a piece of text --- a word, sentence, or passage --- into a fixed-length numerical vector, positioned in a high-dimensional space such that texts with similar meaning are placed close together and texts with dissimilar meaning are placed far apart. Distance between two vectors (commonly cosine similarity) is then used as a proxy for semantic similarity between the corresponding texts. This makes embeddings useful for a wide range of downstream applications: finding the passage that best answers a query, grouping similar documents, detecting duplicate or paraphrased text, and comparing machine-generated text against a reference, among others.

No single embedding model is optimal for all of these applications, because the notion of ``similarity'' a model learns depends entirely on what it was trained to do. A model trained to place paraphrases of the same sentence close together will not necessarily place a question and its correct long-form answer close together, since the two are worded very differently even when one correctly answers the other. Models also differ in scale (parameter count and embedding dimension), the languages they support, the maximum input length they can encode before truncating, and whether they are optimized for a single task or several simultaneously. These differences are the reason a benchmarking study of this kind is needed: choosing an embedding model is not a matter of picking the ``best'' one in the abstract, but the one whose training objective and constraints match the task at hand.

\subsection{Background: The Massive Text Embedding Benchmark (MTEB)}

The Massive Text Embedding Benchmark (MTEB) is a community-maintained, standardized benchmark suite that allows embedding models to be compared on equal footing across a wide range of task types, rather than each model being reported against a different, hand-picked set of datasets. It fixes the datasets, task categories, and evaluation metrics used for each type of task, so that a score obtained by one model is directly comparable to a score obtained by another. Because of this standardization, MTEB has become the de facto reference leaderboard for text embedding models generally, and is widely cited whenever a new embedding model is released \cite{muennighoff2022mteb}. This study adopts MTEB's datasets, model pool, and metrics for the parts of the evaluation that extend beyond the primary retrieval comparison.

\subsection{Research Questions}

Three questions were framed at the outset:

\begin{enumerate}[label=(\arabic*)]
    \item Whether a longer context window and asymmetric query/document encoding provide a measurable quality advantage in retrieval.
    \item How models trained for sentence-similarity tasks (rather than retrieval) perform when repurposed for retrieval.
    \item How sensitive retrieval quality is to the size and method of document chunking.
\end{enumerate}

A fourth, broader question is addressed through the incorporation of MTEB:

\begin{enumerate}[label=(\arabic*), start=4]
    \item Whether models that perform well on retrieval also perform well on other downstream tasks (classification, clustering, semantic similarity, reranking, pair classification, bitext mining, and summarization), or whether task specialization limits generalization.
\end{enumerate}

\section{Background: From Text to Retrieved Results}
\label{sec:background}

Section~\ref{sec:embedding-definition} defined a text embedding at a conceptual level. This section traces the full mechanical pipeline by which a piece of text becomes a retrievable result in a real system, and introduces the infrastructure --- vector databases and approximate nearest neighbor search --- that this study's latency and retrieval-quality figures ultimately depend on.

\begin{figure}[htbp]
    \centering
    \includegraphics[width=\textwidth]{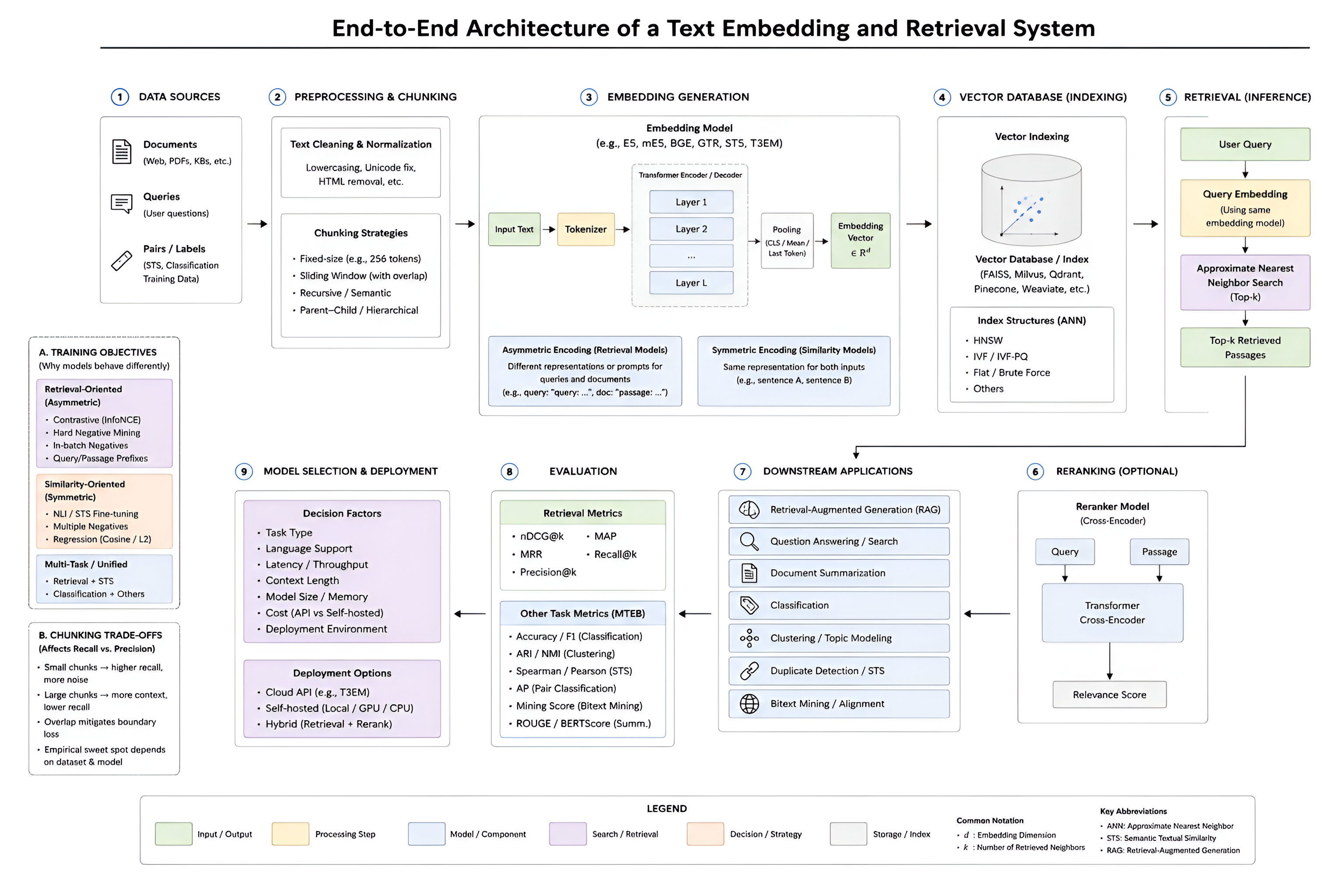}
    \caption{End-to-end architecture of a text embedding and retrieval system, from raw data sources through embedding generation, vector indexing, retrieval, optional reranking, downstream applications, evaluation, and final model selection. Panels A and B summarize, respectively, how differing training objectives (retrieval-oriented, similarity-oriented, or multi-task) produce the behavioral differences discussed throughout this report, and the recall/precision trade-offs introduced by chunk size (Section~2.5). The numbered stages (1--9) correspond broadly to the following sections of this report: data and chunking (Section~\ref{sec:chunking-strategies}), embedding generation (Sections~\ref{sec:how-embeddings-work}--\ref{sec:dense-sparse-multivector}), vector indexing and ANN search (Section~\ref{sec:vector-db-ann}), retrieval and reranking (Section~\ref{sec:vector-db-ann} and Section~\ref{sec:reranking-datasets}), downstream applications and evaluation metrics (Sections~\ref{sec:models}--\ref{sec:metrics}), and model selection (Section~\ref{sec:discussion}).}
    \label{fig:architecture}
\end{figure}

As Figure~\ref{fig:architecture} illustrates, embedding model choice is only one component of a much longer pipeline --- one that also depends on chunking strategy, indexing method, and whether a reranking stage is used. The remainder of this section walks through each stage of this pipeline in turn.

\subsection{How Text Embeddings Work}
\label{sec:how-embeddings-work}

At inference time, a piece of text passes through a fixed sequence of stages before it can be compared against other text:

\begin{center}
\textbf{Transformer} $\rightarrow$ \textbf{Pooling} $\rightarrow$ \textbf{Embedding Vector} $\rightarrow$ \textbf{Cosine Similarity} $\rightarrow$ \textbf{Vector Search} $\rightarrow$ \textbf{ANN Search} $\rightarrow$ \textbf{Retrieved Document}
\end{center}

\begin{enumerate}
    \item \textbf{Transformer encoding.} The input text is tokenized and passed through the model's transformer layers, which use self-attention to produce a contextualized vector for every token --- each token's representation is informed by the tokens around it, not just its own identity.
    \item \textbf{Pooling.} A single fixed-length vector is needed to represent the whole input, so the per-token vectors are combined into one via a pooling strategy: \emph{mean pooling} (averaging all token vectors), \emph{CLS-token pooling} (using a single dedicated summary token's representation), or, less commonly, \emph{max pooling}. The pooling method is typically fixed by how the model was trained and materially affects embedding quality.
    \item \textbf{Embedding vector.} The result is the model's fixed-length numerical representation of the input --- its length is the ``Dim'' figure reported in the model tables in Section~\ref{sec:models}.
    \item \textbf{Cosine similarity.} At query time, the query's embedding is compared against candidate document embeddings using cosine similarity: the cosine of the angle between two vectors, which isolates directional (semantic) alignment while ignoring vector magnitude.
    \item \textbf{Vector search.} In principle, finding the most similar document means comparing the query vector against \emph{every} stored document vector (exact, brute-force search). This is correct but computationally infeasible once a corpus reaches millions of documents.
    \item \textbf{ANN search.} To make search practical at scale, an approximate nearest neighbor (ANN) index is used instead of brute-force comparison, trading a small, controlled amount of accuracy for large gains in speed (detailed in Section~2.3).
    \item \textbf{Retrieved document.} The index returns the top-$k$ most similar documents to the query. In a RAG pipeline, this retrieved text is what gets passed to the language model as context for generating an answer.
\end{enumerate}

\subsection{Dense, Sparse, and Multi-Vector Retrieval}
\label{sec:dense-sparse-multivector}

Retrieval-oriented embedding models fall into three broad architectural approaches, referenced elsewhere in this report (notably in the discussion of BGE-M3's design in Section~\ref{sec:major-tradeoffs}):

\begin{itemize}
    \item \textbf{Dense retrieval} represents each piece of text as a single, fixed-length, densely-packed vector (typically a few hundred to a few thousand dimensions, with almost no zero values). Similarity is computed as a single cosine or dot-product score between two such vectors. This is the approach used by the vast majority of models in this report (T3EM, E5, GTR, and similar).
    \item \textbf{Sparse retrieval} represents each piece of text as a very high-dimensional vector --- one dimension per vocabulary term --- where almost all values are zero and non-zero values indicate term importance (classical term-weighting schemes like BM25~\cite{robertson2009bm25}, or learned variants such as SPLADE~\cite{formal2021splade}, work this way). Sparse retrieval tends to preserve exact keyword matches well, which dense embeddings can sometimes miss.
    \item \textbf{Multi-vector retrieval} represents each piece of text as a \emph{set} of vectors (e.g., one per token) rather than a single vector, and computes similarity via a ``late interaction'' mechanism that compares individual token vectors between query and document before aggregating a score (the approach popularized by ColBERT~\cite{khattab2020colbert}).
\end{itemize}

Models such as BGE-M3 are explicitly trained to support all three modes simultaneously within one model. As discussed in Section~\ref{sec:major-tradeoffs}, this versatility comes at a measurable cost to peak performance on plain dense retrieval compared to a model trained on a single objective.

\subsection{Vector Databases and Approximate Nearest Neighbor Search}
\label{sec:vector-db-ann}

Embeddings alone do not retrieve anything --- a vector database is the infrastructure that stores embeddings alongside their source documents and makes similarity search fast at scale. This is the missing link in the pipeline described above: without it, ``vector search'' would mean comparing a query against every document vector in the corpus one by one, which does not scale beyond a few tens of thousands of documents.

\paragraph{How a vector database works.} A vector database ingests each document's embedding (plus metadata such as the source text and any filters), and builds an \emph{index} --- a data structure organized specifically to make similarity search fast, at the cost of some search accuracy. At query time, the query embedding is compared only against a small, cleverly chosen subset of stored vectors rather than the entire corpus, which is what makes sub-second search over millions or billions of vectors possible.

\paragraph{Approximate Nearest Neighbor (ANN) search.} ``Approximate'' means the returned top-$k$ results are very likely, but not guaranteed, to be the true top-$k$ nearest vectors. This trade-off is deliberate: in exchange for a small, tunable recall loss, ANN indexes are orders of magnitude faster than exact search. The three techniques most commonly underlying ANN indexes are:

\begin{itemize}
    \item \textbf{HNSW (Hierarchical Navigable Small World graphs)~\cite{malkov2018hnsw}.} Organizes vectors into a multi-layered graph where each vector is linked to a small number of its neighbors; search proceeds by ``hopping'' through the graph, starting from a coarse top layer and descending into finer layers. HNSW offers very high recall and speed but consumes considerably more memory than alternatives, since the full graph structure must be held alongside the vectors.
    \item \textbf{IVF (Inverted File Index).} Clusters all vectors into a fixed number of buckets (via k-means), and at query time searches only the buckets nearest to the query rather than the whole dataset. IVF is more memory-efficient than HNSW but requires careful tuning of the number of clusters searched (\texttt{nprobe}) to balance speed against recall.
    \item \textbf{PQ (Product Quantization)~\cite{jegou2011pq}.} Compresses each vector into a compact code by splitting it into sub-vectors and replacing each with the nearest of a small set of representative values. PQ dramatically reduces memory footprint (often by an order of magnitude or more) at the cost of some precision, and is frequently combined with IVF (as ``IVF-PQ'') to get both speed and memory efficiency.
\end{itemize}

\paragraph{Common vector database systems.} The table below summarizes the most widely used systems, their deployment model, and their main limitations.

\begin{longtable}{L{2cm}L{2.8cm}L{5cm}L{4cm}}
\toprule
\textbf{System} & \textbf{Deployment} & \textbf{Notes} & \textbf{Main Drawbacks} \\
\midrule
\endhead
FAISS~\cite{johnson2019faiss} & Embedded library (no server) & Developed by Meta; the reference implementation for most ANN algorithms (HNSW, IVF, PQ); extremely fast and widely used as a building block inside other systems & Not a full database --- no built-in persistence, metadata filtering, or multi-user access; requires custom engineering to productionize \\
Qdrant~\cite{qdrant} & Self-hosted server or managed cloud & Open-source, written in Rust; strong metadata filtering and a straightforward API; good balance of performance and ease of operation & Smaller ecosystem and community than more established options; horizontal scaling is less battle-tested at very large scale \\
Milvus~\cite{wang2021milvus} & Self-hosted server or managed cloud & Open-source; built for very large-scale deployments with distributed architecture and support for multiple index types & Operationally heavier to run and tune than Qdrant; a distributed system introduces real infrastructure overhead for smaller deployments \\
Pinecone~\cite{pinecone} & Fully managed cloud only & Proprietary, fully managed service; minimal operational overhead, scales transparently & Not open-source; ongoing usage cost; no self-hosting option, so data resides in a third-party managed environment \\
\bottomrule
\end{longtable}

Among open-source options, \textbf{Qdrant} is generally the preferred starting point for small to mid-scale deployments due to its balance of performance, ease of self-hosting, and filtering support, while \textbf{Milvus} is typically preferred once scale (hundreds of millions of vectors or more) or distributed deployment becomes a hard requirement. \textbf{FAISS} remains the standard choice when a lightweight, embedded solution is sufficient and a full database server is unnecessary overhead.

\paragraph{The full retrieval pipeline.} Putting the embedding and infrastructure pieces together, a typical retrieval-augmented system follows this sequence:

\begin{center}
\textbf{Query} $\rightarrow$ \textbf{Embedding} $\rightarrow$ \textbf{Vector DB} $\rightarrow$ \textbf{ANN Search} $\rightarrow$ \textbf{Top-\emph{k}} $\rightarrow$ \textbf{Reranker} $\rightarrow$ \textbf{LLM}
\end{center}

The reranking stage (see the Reranking Datasets category in Section~\ref{sec:reranking-datasets}) is typically applied after the initial ANN search, since a small, more expensive model can afford to carefully re-score a short list of already-retrieved candidates in a way that would be too slow to apply against the full corpus directly.

\subsection{Document Chunking Strategies}
\label{sec:chunking-strategies}

Because embedding models truncate input beyond their maximum context length (Section~3), long documents must be split into smaller pieces --- \emph{chunks} --- before being embedded and indexed. The choice of chunking strategy has a direct effect on retrieval quality (empirically measured in Section~6.3) and is one of the most consequential design decisions in a retrieval pipeline.

\begin{itemize}
    \item \textbf{Fixed-size chunking.} Splits text into chunks of a constant token or character length, regardless of sentence or paragraph boundaries. Simple and fast to implement, but can cut sentences or ideas in half, splitting relevant information across two chunks.
    \item \textbf{Sliding window chunking.} Similar to fixed-size chunking, but consecutive chunks overlap by a fixed amount rather than starting exactly where the previous one ended. This reduces the chance that a relevant passage is split awkwardly across a chunk boundary, at the cost of some redundant storage.
    \item \textbf{Semantic chunking.} Splits text at natural topic or meaning boundaries (e.g., where embedding similarity between consecutive sentences drops sharply) rather than at a fixed length. This tends to produce more coherent, self-contained chunks, though it is more computationally expensive to compute than fixed-size splitting.
    \item \textbf{Parent-child chunking.} Indexes small, precise chunks for matching (the ``children''), but retrieves and passes the larger surrounding section (the ``parent'') to the generation stage once a child chunk matches. This combines precise retrieval matching with sufficient surrounding context for the language model to use.
    \item \textbf{Hierarchical chunking.} Builds multiple levels of chunks --- e.g., document-level, section-level, and paragraph-level --- and allows retrieval to operate at whichever level is most appropriate for a given query, or to combine evidence across levels.
    \item \textbf{Recursive chunking.} Splits text using a prioritized list of separators (e.g., first by paragraph breaks, then by sentence breaks, then by word breaks only if a piece is still too long), aiming to keep each chunk as semantically intact as possible while still respecting a maximum length.
    \item \textbf{Chunk overlap.} A parameter (rather than a strategy on its own) that can be applied to most of the above: a fixed number of tokens repeated between consecutive chunks, so that context near a chunk boundary is not entirely lost to either chunk alone.
\end{itemize}

\paragraph{When to use each.} Fixed-size and sliding-window chunking are reasonable defaults for homogeneous, well-structured text where sentence-level precision is less critical. Semantic chunking is preferable when documents cover multiple distinct topics within a single file (e.g., long-form articles or manuals) and retrieval precision matters more than implementation simplicity. Parent-child and hierarchical chunking are best suited to long, structured documents (legal contracts, technical documentation) where a precise passage needs to be matched but a wider surrounding context is required for the answer to make sense. Recursive chunking is a practical general-purpose default in most modern RAG frameworks, since it approximates semantic boundaries without the computational cost of true semantic chunking. As found in this study (Section~6.3), gains from chunking strategy diminish sharply above 32 tokens per chunk, and all strategies degrade below roughly 16 tokens, regardless of which method is used.

\section{Datasets}
\label{sec:datasets}

All datasets used across the retrieval evaluation and the wider MTEB framework are consolidated below, organized by task category. Four of the fifteen retrieval datasets (FiQA-2018, NFCorpus, SciFact, and TREC-COVID) were used for the primary, directly measured retrieval comparison in this study (Section~\ref{sec:results}); the remaining datasets are drawn from the published MTEB framework and are referenced for broader context rather than re-measured here. This distinction is indicated in the \textit{Primary Evaluation} column of Table~\ref{tab:retrieval-datasets}.

\subsection{Retrieval Datasets}

These test a model's ability to find the right passage or document out of a large collection, given a query. This is the core task behind search engines and RAG (retrieval-augmented generation) systems. The query and the correct answer are often worded very differently (e.g., a question vs. a passage that answers it), so retrieval datasets specifically test whether a model can match meaning rather than just overlapping words. For example, a FiQA-style query such as ``Should I switch to a Roth IRA?'' is paired with a financial forum passage discussing tax treatment using vocabulary quite different from that of the question itself, making it a useful stress test for asymmetric query/document encoding.

\begin{longtable}{L{2.6cm}L{3.8cm}L{2.2cm}L{5.2cm}}
\caption{Retrieval datasets used in this study.\label{tab:retrieval-datasets}}\\
\toprule
\textbf{Dataset} & \textbf{Description} & \textbf{Primary Eval.} & \textbf{Link} \\
\midrule
\endfirsthead
\toprule
\textbf{Dataset} & \textbf{Description} & \textbf{Primary Eval.} & \textbf{Link} \\
\midrule
\endhead
FiQA-2018 & Financial question answering retrieval (648 queries) & Yes & \url{https://huggingface.co/datasets/mteb/fiqa} \\
NFCorpus & Biomedical retrieval (323 queries) & Yes & \url{https://huggingface.co/datasets/mteb/nfcorpus} \\
SciFact & Scientific claim / evidence verification retrieval (300 queries) & Yes & \url{https://huggingface.co/datasets/mteb/scifact} \\
TREC-COVID & Biomedical / COVID literature retrieval (50 queries) & Yes & \url{https://huggingface.co/datasets/mteb/trec-covid} \\
ArguAna & Counter-argument retrieval & No (MTEB) & \url{https://huggingface.co/datasets/mteb/arguana} \\
ClimateFEVER & Climate fact verification retrieval & No (MTEB) & \url{https://huggingface.co/datasets/mteb/climate-fever} \\
CQADupStack & Technical question retrieval & No (MTEB) & \url{https://huggingface.co/datasets/mteb/cqadupstack-android} \\
DBPedia & Entity retrieval & No (MTEB) & \url{https://huggingface.co/datasets/mteb/dbpedia} \\
FEVER & Evidence retrieval & No (MTEB) & \url{https://huggingface.co/datasets/mteb/fever} \\
HotpotQA & Multi-hop retrieval & No (MTEB) & \url{https://huggingface.co/datasets/mteb/hotpotqa} \\
MSMARCO & Web search passage retrieval & No (MTEB) & \url{https://huggingface.co/datasets/mteb/msmarco} \\
Natural Questions & Google search retrieval & No (MTEB) & \url{https://huggingface.co/datasets/mteb/nq} \\
Quora Retrieval & Duplicate question retrieval & No (MTEB) & \url{https://huggingface.co/datasets/mteb/quora} \\
SciDocs & Scientific paper retrieval & No (MTEB) & \url{https://huggingface.co/datasets/mteb/scidocs} \\
Touche2020 & Argument retrieval & No (MTEB) & \url{https://huggingface.co/datasets/mteb/touche2020} \\
\bottomrule
\end{longtable}

\subsection{Semantic Textual Similarity (STS) Datasets}

These measure whether a model can judge how similar in meaning two sentences are, on a continuous scale (rather than a binary yes/no). Human annotators score sentence pairs for similarity, and the model's job is to produce embeddings whose distances correlate with those human judgments. This is a symmetric task — unlike retrieval, both sentences are of the same "type" (e.g., two statements), not a short query against a long document.

\begin{longtable}{L{3cm}L{4cm}L{6cm}}
\toprule
\textbf{Dataset} & \textbf{Description} & \textbf{Link} \\
\midrule
\endhead
BIOSSES & Biomedical sentence similarity & \url{https://huggingface.co/datasets/mteb/biosses-sts} \\
SICK-R & Sentence similarity & \url{https://huggingface.co/datasets/mteb/sickr-sts} \\
STS12 & SemEval STS & \url{https://huggingface.co/datasets/mteb/sts12-sts} \\
STS13 & SemEval STS & (same naming convention as STS12) \\
STS14 & SemEval STS & (same naming convention as STS12) \\
STS15 & SemEval STS & (same naming convention as STS12) \\
STS16 & SemEval STS & (same naming convention as STS12) \\
STS17 & Cross-lingual STS & \url{https://huggingface.co/datasets/mteb/sts17-crosslingual-sts} \\
STS22 & Multilingual STS & \url{https://huggingface.co/datasets/mteb/sts22-crosslingual-sts} \\
STSBenchmark & Standard English STS & \url{https://huggingface.co/datasets/mteb/stsbenchmark-sts} \\
\bottomrule
\end{longtable}

\subsection{Classification Datasets}

These assign a single category label to a piece of text — sentiment (positive/negative), intent (e.g., banking query type), topic, or similar. The embedding model isn't directly trained to classify; instead, its embeddings are typically fed into a simple classifier (like logistic regression), and the benchmark checks whether the embeddings alone carry enough signal to separate the categories well.

\begin{longtable}{L{3cm}L{4cm}L{6cm}}
\toprule
\textbf{Dataset} & \textbf{Description} & \textbf{Link} \\
\midrule
\endhead
Amazon Counterfactual & Counterfactual review detection & \url{https://huggingface.co/datasets/mteb/amazon_counterfactual} \\
Amazon Polarity & Sentiment classification & \url{https://huggingface.co/datasets/mteb/amazon_polarity} \\
Amazon Reviews & Review rating prediction & \url{https://huggingface.co/datasets/mteb/amazon_reviews_multi} \\
Banking77 & Banking intent classification & \url{https://huggingface.co/datasets/mteb/banking77} \\
Emotion & Emotion recognition & \url{https://huggingface.co/datasets/mteb/emotion} \\
IMDb & Movie review sentiment & \url{https://huggingface.co/datasets/mteb/imdb} \\
Massive Intent & Intent classification & \url{https://huggingface.co/datasets/mteb/amazon_massive_intent} \\
Massive Scenario & Scenario classification & \url{https://huggingface.co/datasets/mteb/amazon_massive_scenario} \\
MTOP Domain & Dialogue domain classification & \url{https://huggingface.co/datasets/mteb/mtop_domain} \\
Toxic Conversations & Toxicity detection & \url{https://huggingface.co/datasets/mteb/toxic_conversations_50k} \\
Tweet Sentiment & Tweet sentiment classification & \url{https://huggingface.co/datasets/mteb/tweet_sentiment_extraction} \\
\bottomrule
\end{longtable}

\subsection{Clustering Datasets}

These test whether embeddings naturally group similar documents together without being told the categories in advance (unsupervised). For example, scientific papers should cluster by subfield if the embeddings capture topical meaning well. This is a useful proxy for how well a model organizes a large, unlabeled corpus.

\begin{longtable}{L{3cm}L{4cm}L{6cm}}
\toprule
\textbf{Dataset} & \textbf{Description} & \textbf{Link} \\
\midrule
\endhead
Arxiv & Scientific paper clustering & \url{https://huggingface.co/datasets/mteb/arxiv-clustering-p2p} \\
BioRxiv & Biological paper clustering & \url{https://huggingface.co/datasets/mteb/biorxiv-clustering-p2p} \\
MedRxiv & Medical paper clustering & \url{https://huggingface.co/datasets/mteb/medrxiv-clustering-p2p} \\
Reddit & Reddit post clustering & \url{https://huggingface.co/datasets/mteb/reddit-clustering} \\
StackExchange & Technical forum clustering & \url{https://huggingface.co/datasets/mteb/stackexchange-clustering} \\
TwentyNewsgroups & News article clustering & \url{https://huggingface.co/datasets/mteb/twentynewsgroups-clustering} \\
\bottomrule
\end{longtable}

\subsection{Pair Classification Datasets}

These are a specific yes/no version of similarity: given two pieces of text, is this pair a duplicate/paraphrase or not? It's similar in spirit to STS but framed as a binary decision (duplicate vs. not) rather than a graded similarity score — common in deduplication and spam/near-duplicate detection use cases.

\begin{longtable}{L{3cm}L{3.5cm}L{7cm}}
\toprule
\textbf{Dataset} & \textbf{Description} & \textbf{Link} \\
\midrule
\endhead
Sprint Duplicate Questions & Duplicate detection & \url{https://huggingface.co/datasets/mteb/sprintduplicatequestions-pairclassification} \\
Twitter SemEval & Tweet paraphrase detection & \url{https://huggingface.co/datasets/mteb/twittersemeval2015-pairclassification} \\
Twitter URL Corpus & URL paraphrase detection & \url{https://huggingface.co/datasets/mteb/twitterurlcorpus-pairclassification} \\
\bottomrule
\end{longtable}

\subsection{Reranking Datasets}
\label{sec:reranking-datasets}

These test a second-stage task common in real search pipelines: given a query and a list of candidate documents (already loosely retrieved), can the model correctly reorder them so the most relevant ones rise to the top? It's a sharper, more precision-focused version of retrieval, usually applied after an initial, cheaper retrieval pass.

\begin{longtable}{L{3.5cm}L{4cm}L{6cm}}
\toprule
\textbf{Dataset} & \textbf{Description} & \textbf{Link} \\
\midrule
\endhead
AskUbuntu & Ubuntu question reranking & \url{https://huggingface.co/datasets/mteb/askubuntudupquestions-reranking} \\
MindSmall & News reranking & \url{https://huggingface.co/datasets/mteb/mind_small} \\
SciDocsRR & Scientific reranking & \url{https://huggingface.co/datasets/mteb/scidocs-reranking} \\
StackOverflow & Duplicate question reranking & \url{https://huggingface.co/datasets/mteb/stackoverflowdupquestions-reranking} \\
\bottomrule
\end{longtable}

\subsection{Bitext Mining Datasets}

These test cross-lingual alignment — given a sentence in one language, can the model find its true translation in a pool of sentences in another language? This is the main way multilingual sentence-alignment quality is evaluated, and it's central to building parallel corpora for machine translation.

\begin{longtable}{L{3cm}L{4cm}L{6cm}}
\toprule
\textbf{Dataset} & \textbf{Description} & \textbf{Link} \\
\midrule
\endhead
BUCC & Parallel sentence mining & \url{https://huggingface.co/datasets/mteb/bucc-bitext-mining} \\
Tatoeba & Translation mining & \url{https://huggingface.co/datasets/mteb/tatoeba-bitext-mining} \\
\bottomrule
\end{longtable}

\subsection{Summarization Datasets}

These evaluate whether a model's embeddings can judge the quality of an automatically generated summary by comparing it (via embedding similarity) to reference human summaries. It's a proxy for whether embedding-based similarity correlates with human judgments of summary quality — rather than the model generating summaries itself.

\begin{longtable}{L{3cm}L{4cm}L{6cm}}
\toprule
\textbf{Dataset} & \textbf{Description} & \textbf{Link} \\
\midrule
\endhead
SummEval & Automatic summary evaluation & \url{https://huggingface.co/datasets/mteb/summeval} \\
\bottomrule
\end{longtable}

\section{Models Evaluated}
\label{sec:models}

All models referenced in this study are consolidated below. Models are grouped first by whether their retrieval performance was directly measured for this study's primary evaluation, and then by architectural family for the wider set of models drawn from the MTEB framework and from published leaderboard estimates. Grouping reflects model type rather than any distinction between separate studies.

\paragraph{Note on context length.} Every embedding model has a maximum input length (measured in tokens) that it can process in a single pass. If an input document exceeds this limit, the excess text is silently truncated --- that is, discarded without warning --- before the embedding is computed. For a model with a 512-token limit, this means only roughly the first few hundred words of a long document are actually reflected in its embedding, regardless of how much additional relevant content follows. This is why document length and chunking strategy (Sections~\ref{sec:chunking-strategies} and \ref{sec:chunking-results}) become important practical considerations when selecting a model for long-document retrieval.

\subsection{Models with Directly Measured Retrieval Results}
\label{sec:primary-models}

The six models below were evaluated directly on the primary English retrieval benchmark subsets described in Section~\ref{sec:datasets} (Table~\ref{tab:retrieval-datasets}; results in Section~\ref{sec:results}).

\begin{longtable}{L{3.4cm}L{2.3cm}L{2cm}L{6.3cm}}
\toprule
\textbf{Model} & \textbf{Type} & \textbf{Open Source} & \textbf{Link} \\
\midrule
\endhead
T3EM (Text 3 Embedding Model) & API & No (commercial API) & --- \\
BGE-M3~\cite{chen2024bgem3} & Open-source & Yes & \url{https://huggingface.co/BAAI/bge-m3} \\
E5-large~\cite{wang2022e5} & Open-source & Yes & \url{https://huggingface.co/intfloat/e5-large-v2} \\
Multilingual-E5-large (mE5-L)~\cite{wang2024multilinguale5} & Open-source & Yes & \url{https://huggingface.co/intfloat/multilingual-e5-large} \\
LaBSE~\cite{feng2022labse} & Open-source & Yes & \url{https://huggingface.co/sentence-transformers/LaBSE} \\
Paraphrase-Multilingual-MPNet (mMPNet)~\cite{reimers2019sbert,song2020mpnet} & Open-source & Yes & \url{https://huggingface.co/sentence-transformers/paraphrase-multilingual-mpnet-base-v2} \\
\bottomrule
\end{longtable}

\subsection{Additional Models Evaluated Under the MTEB Framework}

The models below extend the comparison to a much larger pool evaluated under the published MTEB framework, grouped by architectural family. Their scores (Section~\ref{sec:results}) are drawn from MTEB rather than re-measured in this study's own retrieval pipeline. Note that LaBSE and Paraphrase-Multilingual-MPNet already appear in Section~\ref{sec:primary-models} above and are not repeated here.

\subsubsection{Self-Supervised Models}

These are trained without labeled data, purely on raw text, using generic objectives like predicting missing or nearby words. They weren't designed specifically for retrieval or similarity — they're general-purpose language representations (e.g., GloVe~\cite{pennington2014glove}, Komninos embeddings~\cite{komninos2016dependency}, BERT~\cite{devlin2019bert}) that serve as a baseline for how much task-specific fine-tuning actually helps.

\begin{longtable}{L{4cm}cL{7cm}}
\toprule
\textbf{Model} & \textbf{Open Source} & \textbf{Link} \\
\midrule
\endhead
GloVe & Yes & \url{https://huggingface.co/sentence-transformers/average_word_embeddings_glove.6B.300d} \\
Komninos & Yes & \url{https://huggingface.co/sentence-transformers/average_word_embeddings_komninos} \\
BERT Base & Yes & \url{https://huggingface.co/bert-base-uncased} \\
\bottomrule
\end{longtable}

\subsubsection{Contrastively Fine-Tuned Models}

These start from a pretrained language model and are further trained using a "contrastive" objective: pulling embeddings of similar/paired texts closer together and pushing dissimilar ones apart. This is the dominant recipe behind most modern sentence and retrieval embedding models (e.g., SimCSE~\cite{gao2021simcse}, Contriever~\cite{izacard2022contriever}, SPECTER~\cite{cohan2020specter}, MPNet~\cite{song2020mpnet}, built via the Sentence-BERT framework~\cite{reimers2019sbert}) because it directly optimizes for the property embeddings actually need — meaningful distances.

\begin{longtable}{L{4cm}cL{7cm}}
\toprule
\textbf{Model} & \textbf{Open Source} & \textbf{Link} \\
\midrule
\endhead
SimCSE & Yes & \url{https://huggingface.co/princeton-nlp/sup-simcse-bert-base-uncased} \\
coCondenser & Yes & \url{https://huggingface.co/sentence-transformers/msmarco-bert-co-condensor} \\
Contriever & Yes & \url{https://huggingface.co/nthakur/contriever-base-msmarco} \\
SPECTER & Yes & \url{https://huggingface.co/sentence-transformers/allenai-specter} \\
MiniLM (all-MiniLM-L12-v2) & Yes & \url{https://huggingface.co/sentence-transformers/all-MiniLM-L12-v2} \\
MPNet (all-mpnet-base-v2, English) & Yes & \url{https://huggingface.co/sentence-transformers/all-mpnet-base-v2} \\
\bottomrule
\end{longtable}

\subsubsection{T5 Encoder-Based Models}

These repurpose the encoder half of T5 (a text-to-text transformer originally built for generation) as a pure embedding generator, then fine-tune it on similarity or retrieval objectives. Because T5 was pretrained at large scale for general language understanding, these models (e.g., GTR~\cite{ni2021gtr}, ST5~\cite{ni2021st5}) tend to produce strong, well-rounded embeddings across many tasks.

\begin{longtable}{L{4cm}cL{7cm}}
\toprule
\textbf{Model} & \textbf{Open Source} & \textbf{Link} \\
\midrule
\endhead
GTR & Yes & \url{https://huggingface.co/sentence-transformers/gtr-t5-xxl} \\
ST5 & Yes & \url{https://huggingface.co/sentence-transformers/sentence-t5-xxl} \\
\bottomrule
\end{longtable}

\subsubsection{Decoder-Based Models}

These derive embeddings from decoder-only, GPT-style language models (e.g., SGPT) rather than encoder architectures. Historically decoders were seen as less natural for embeddings (since they're built for next-word generation, not bidirectional understanding), so this family tests whether large generative models can be adapted into competitive embedding models simply by pooling their internal representations \cite{muennighoff2022sgpt}.

\begin{longtable}{L{4cm}cL{7cm}}
\toprule
\textbf{Model} & \textbf{Open Source} & \textbf{Link} \\
\midrule
\endhead
SGPT & Yes & \url{https://huggingface.co/Muennighoff/SGPT-5.8B-weightedmean-msmarco-specb-bitfit} \\
SGPT BLOOM & Yes & \url{https://huggingface.co/bigscience/sgpt-bloom-7b1-msmarco} \\
\bottomrule
\end{longtable}

\subsubsection{Additional Multilingual-Capable Models}

These are models specifically trained or fine-tuned to represent text across many languages in a shared embedding space, rather than being optimized for a single language (e.g., LaBSE, LASER2, multilingual MiniLM). Their main value is enabling cross-lingual tasks — like bitext mining or multilingual retrieval — usually at some cost to peak performance on any single language compared to a monolingual specialist. LASER2 builds on the original LASER architecture \cite{artetxe2019laser}.

\begin{longtable}{L{4.5cm}cL{6.5cm}}
\toprule
\textbf{Model} & \textbf{Open Source} & \textbf{Link} \\
\midrule
\endhead
LASER2 & Yes & \url{https://github.com/facebookresearch/LASER} \\
MiniLM Multilingual & Yes & \url{https://huggingface.co/sentence-transformers/paraphrase-multilingual-MiniLM-L12-v2} \\
\bottomrule
\end{longtable}

\subsubsection{Closed-Source Model (MTEB)}

\begin{table}[h]
\centering
\begin{tabular}{lll}
\toprule
\textbf{Model} & \textbf{Type} & \textbf{Link} \\
\midrule
OpenAI Ada Similarity / Ada Search & Commercial API & Not open source \\
\bottomrule
\end{tabular}
\end{table}

\subsection{Additional Open-Source Models (Estimated Performance Only)}
\label{sec:estimated-models}

The models below were not part of the primary benchmark or the MTEB evaluation. The figures reported for them in Section~\ref{sec:results} are estimated scores based on published retrieval performance (e.g.\ MTEB leaderboard standing) rather than measurements taken under this study's own evaluation pipeline, and are presented separately so as not to be confused with directly measured results.

\paragraph{Note on table columns.} In the tables below, \textbf{Dim} refers to the embedding dimension --- the length of the output vector produced by the model --- which directly affects how much storage and memory is required to index a large corpus. \textbf{Params} refers to the total number of trainable parameters in the model, which is a rough proxy for its computational cost at inference time: larger models generally require more memory and take longer to run.

\begin{longtable}{L{3.4cm}ccL{2.6cm}L{4cm}}
\toprule
\textbf{Model} & \textbf{Dim} & \textbf{Params} & \textbf{Est. BEIR-style nDCG@10} & \textbf{Notes} \\
\midrule
\endhead
all-MiniLM-L6-v2 & 384 & 22M & $\sim$0.42--0.45 & Classic lightweight SBERT baseline; fast but dated for retrieval \\
all-MiniLM-L12-v2 & 384 & 33M & $\sim$0.44--0.47 & Slightly deeper than L6, marginal quality gain \\
Nomic-Embed-Text-v1.5~\cite{nussbaum2024nomic} & 768 & 137M & $\sim$0.55--0.58 & Strong for its size; long-context (8192 tokens) capable \\
BGE-base-en-v1.5~\cite{xiao2023bge} & 768 & 109M & $\sim$0.53--0.56 & Standard RAG baseline, English-only \\
BGE-large-en-v1.5~\cite{xiao2023bge} & 1024 & 335M & $\sim$0.57--0.60 & One of the strongest English-only open baselines, close to T3EM on English tasks \\
Qwen3-Embedding-0.6B & 1024 & 600M & $\sim$0.58--0.61 & Newer generation; competitive retrieval at small size \\
Qwen3-Embedding-4B & 2560 & 4B & $\sim$0.63--0.66 & Largest model here; expected to approach or match T3EM on BEIR average \\
\bottomrule
\end{longtable}

\section{Evaluation Metrics}
\label{sec:metrics}

\subsection{Symmetric vs.\ Asymmetric Embedding Tasks}

Embedding tasks can be broadly divided into two types, and this distinction underlies much of the model behavior discussed later in this report.

In a \textbf{symmetric} task, the two pieces of text being compared are of similar length, structure, and purpose --- for example, two full sentences being checked for paraphrase or similarity (as in STS or pair classification). In an \textbf{asymmetric} task, the two pieces of text differ substantially in length, structure, and wording --- most commonly, a short query being matched against a long passage that answers it (as in retrieval). A query rarely repeats the vocabulary of the passage that answers it, so an asymmetric task requires the model to bridge a wording gap that a symmetric task does not.

Models trained primarily on symmetric objectives (e.g.\ paraphrase or sentence-similarity training) do not automatically transfer well to asymmetric retrieval, since they were never trained to bridge that gap. This distinction explains why some models evaluated in this study perform respectably on similarity-style benchmarks but comparatively poorly on retrieval, despite being trained on large amounts of data.

\subsection{Metrics per Task Category}

Each downstream task is evaluated using metrics best suited to its objective.

\begin{table}[h]
\centering
\begin{tabular}{lll}
\toprule
\textbf{Task} & \textbf{Primary Metric} & \textbf{Additional Metrics} \\
\midrule
Retrieval & nDCG@10 & Recall@k, MRR@k, MAP@k, Precision@k \\
Semantic Textual Similarity & Spearman Correlation & Pearson Correlation \\
Classification & Accuracy & F1 Score, Average Precision \\
Clustering & V-Measure & --- \\
Pair Classification & Average Precision & Accuracy, Precision, Recall, F1 \\
Reranking & MAP & MRR@k \\
Bitext Mining & F1 Score & Accuracy, Precision, Recall \\
Summarization & Spearman Correlation & Pearson Correlation \\
\bottomrule
\end{tabular}
\caption{Primary and additional evaluation metrics used per task category.}
\end{table}

\subsection{Interpretation of Metrics}

For retrieval specifically, three metrics recur throughout this report:

\begin{itemize}
    \item \textbf{Recall@k} --- the proportion of relevant passages that appear anywhere within the top-$k$ retrieved results. This indicates how much useful information was surfaced at all, regardless of exact ranking, and is the most directly relevant metric for retrieval-augmented generation (RAG), since a generator can only use what has been retrieved.
    \item \textbf{Mean Reciprocal Rank (MRR)} --- the average of one divided by the rank position of the first correct result. This metric rewards placing the correct passage at rank 1 specifically, and penalizes models that only find the right passage lower down the list.
    \item \textbf{nDCG@10} --- the primary metric used throughout this study. It rewards relevant passages more when they appear near the top of the ranked list and less when they appear further down, using a logarithmic discount. This is the standard metric on BEIR and MTEB retrieval tasks, and was treated as the main quality indicator since it allows direct comparability with other published work.
\end{itemize}

The remaining metrics, used for non-retrieval task categories, are summarized below.

\begin{longtable}{L{3.5cm}L{9.5cm}}
\toprule
\textbf{Metric} & \textbf{Intuitive Meaning} \\
\midrule
\endhead
MAP & Evaluates ranking quality across all relevant documents. \\
Spearman Correlation & Measures agreement between the model's similarity rankings and human judgments. \\
Pearson Correlation & Measures linear correlation between predicted and human similarity scores. \\
Accuracy & Percentage of correctly classified samples. \\
F1 Score & Harmonic mean of precision and recall, balancing false positives and false negatives. \\
Average Precision & Measures ranking quality by rewarding correct predictions that appear earlier in the ranked list. \\
V-Measure & Evaluates clustering quality by measuring cluster homogeneity and completeness. \\
\bottomrule
\end{longtable}

\subsection{Formal Definitions of Metrics}

\paragraph{Discounted Cumulative Gain (DCG) and nDCG@10~\cite{jarvelin2002ndcg}.}
For a ranked list of results,relevance scores $rel_i$ at rank $i$ contribute to the Discounted Cumulative Gain as:
\[
\text{DCG@}k = \sum_{i=1}^{k} \frac{rel_i}{\log_2(i+1)}
\]
The Ideal DCG (IDCG@$k$) is the DCG obtained if results were ranked in the best possible order. Normalized DCG is then:
\[
\text{nDCG@}k = \frac{\text{DCG@}k}{\text{IDCG@}k}
\]
so that a perfect ranking scores 1.0 regardless of how many relevant items exist.

\paragraph{Recall@k.}
\[
\text{Recall@}k = \frac{|\{\text{relevant items}\} \cap \{\text{top-}k\text{ retrieved items}\}|}{|\{\text{relevant items}\}|}
\]

\paragraph{Mean Reciprocal Rank (MRR).}
\[
\text{MRR} = \frac{1}{|Q|} \sum_{i=1}^{|Q|} \frac{1}{\text{rank}_i}
\]
where $|Q|$ is the number of queries and $\text{rank}_i$ is the rank position of the first relevant result for query $i$.

\paragraph{Mean Average Precision (MAP).}
\[
\text{AP} = \frac{\sum_{k=1}^{n} P(k)\cdot \text{rel}(k)}{\text{number of relevant items}}, \qquad
\text{MAP} = \frac{1}{|Q|}\sum_{i=1}^{|Q|} \text{AP}_i
\]
where $P(k)$ is precision at cutoff $k$ and $\text{rel}(k)$ is an indicator that the item at rank $k$ is relevant.

\paragraph{Spearman's $\rho$ and Pearson's $r$.}
Pearson's $r$ measures linear correlation between predicted similarity scores $x_i$ and human judgment scores $y_i$:
\[
r = \frac{\sum_i (x_i - \bar{x})(y_i - \bar{y})}{\sqrt{\sum_i (x_i-\bar{x})^2}\sqrt{\sum_i (y_i-\bar{y})^2}}
\]
Spearman's $\rho$ applies the same formula to the \emph{ranks} of $x_i$ and $y_i$ rather than their raw values, making it robust to non-linear but monotonic relationships.

\paragraph{Precision, Recall, and F1.}
\[
\text{Precision} = \frac{TP}{TP + FP}, \qquad
\text{Recall} = \frac{TP}{TP + FN}, \qquad
F_1 = 2\cdot\frac{\text{Precision}\cdot\text{Recall}}{\text{Precision}+\text{Recall}}
\]
where $TP$, $FP$, and $FN$ denote true positives, false positives, and false negatives respectively.

\paragraph{V-Measure.}
V-Measure is the harmonic mean of homogeneity $h$ (each cluster contains only members of a single class) and completeness $c$ (all members of a class are assigned to the same cluster):
\[
V = \frac{(1+\beta)\cdot h \cdot c}{\beta\cdot h + c}
\]
with $\beta=1$ typically used to weight homogeneity and completeness equally.

\section{Results}
\label{sec:results}

\subsection{Retrieval Performance on Primary Benchmark Subsets}

\begin{table}[h]
\centering
\begin{tabular}{lccccc}
\toprule
\textbf{Model} & \textbf{FiQA} & \textbf{NFCorpus} & \textbf{SciFact} & \textbf{TREC-COVID} & \textbf{Average} \\
\midrule
T3EM      & 0.582 & 0.409 & 0.762 & 0.799 & 0.638 \\
mE5-L     & 0.438 & 0.341 & 0.704 & 0.702 & 0.546 \\
E5-large  & 0.411 & 0.374 & 0.722 & 0.646 & 0.538 \\
BGE-M3    & 0.366 & 0.294 & 0.650 & ---   & 0.437 \\
mMPNet    & 0.174 & 0.172 & 0.317 & 0.308 & 0.243 \\
LaBSE     & 0.069 & 0.155 & 0.378 & 0.151 & 0.188 \\
\bottomrule
\end{tabular}
\caption{nDCG@10 scores on the four primary English BEIR subsets.\label{tab:primary-results}}
\end{table}

\subsection{Query Latency and Cost}
\label{sec:latency-results}

\paragraph{Note on latency figures.} \textbf{Median} latency reflects the typical response time for a query, while \textbf{p95} latency reflects the 95th-percentile response time --- i.e.\ the slowest response experienced by 1 in 20 queries. p95 is generally the more important figure for user-facing applications with latency guarantees (service-level agreements), since it captures worst-case rather than average behavior. For context, T3EM's cost of \$0.025 per 1 million tokens means that embedding roughly 700,000 words of text (a small-to-medium document collection) costs approximately \$0.025, making the per-query cost negligible in isolation but potentially significant at the scale of millions of queries or a large, frequently re-indexed corpus.

\begin{table}[h]
\centering
\begin{tabular}{lccc}
\toprule
\textbf{Model} & \textbf{Median (ms)} & \textbf{p95 (ms)} & \textbf{Cost} \\
\midrule
T3EM     & 231.6 & 575.5 & \$0.025 / 1M tokens \\
BGE-M3   & 30.9  & 32.1  & Free \\
E5-large & 30.9  & 31.4  & Free \\
mE5-L    & 31.0  & 31.8  & Free \\
LaBSE    & 16.6  & 16.9  & Free \\
mMPNet   & 16.6  & 17.0  & Free \\
\bottomrule
\end{tabular}
\caption{Median and 95th-percentile (p95) query latency, and per-query cost.}
\end{table}

\subsection{Effect of Document Chunking Strategy}
\label{sec:chunking-results}

All six primary models reached at least 95\% of their peak nDCG@10 by a chunk size of 32 tokens, with no further gains observed at 64 or 128 tokens.

At a chunk size of 16 tokens, semantic chunking (splitting text at natural topic boundaries rather than at a fixed length) was found to outperform fixed-size chunking by 0.090 nDCG points for T3EM and 0.075 points for mE5-L. Below 16 tokens, quality collapsed for all models, indicating that chunks that small no longer preserve enough coherent meaning to be usefully embedded.

\subsection{Estimated Performance of Additional Open-Source Models}

The estimated scores for the additional models introduced in Section~\ref{sec:estimated-models} are summarized in that section's table and are not repeated here.
 As noted there, these figures are drawn from published leaderboard standings rather than measured directly under this study's retrieval pipeline, and Qwen3-Embedding-4B in particular is expected to approach or match T3EM on BEIR-style averages given its scale.

\subsection{Broader MTEB Performance Summary}
\label{sec:mteb-summary}

\begin{longtable}{L{3.5cm}L{5.5cm}c}
\toprule
\textbf{Model Family} & \textbf{Model} & \textbf{Overall Average Score} \\
\midrule
\endhead
Self-Supervised & GloVe & 41.97 \\
Self-Supervised & Komninos & 42.06 \\
Self-Supervised & BERT & 38.33 \\
Contrastive & SimCSE-BERT (Unsupervised) & 45.45 \\
Contrastive & SimCSE-BERT (Supervised) & 48.72 \\
Contrastive & coCondenser-msmarco & 52.35 \\
Contrastive & Contriever & 56.00 \\
Scientific & SPECTER & 40.28 \\
Multilingual & LaBSE & 45.21 \\
Multilingual & LASER2 & 34.95 \\
Sentence Transformer & MiniLM-L6 & 56.26 \\
Sentence Transformer & MiniLM-L12 & 56.53 \\
Multilingual & MiniLM-L12 Multilingual & 52.44 \\
Sentence Transformer & MPNet & 57.78 \\
Multilingual & MPNet Multilingual & 54.71 \\
Closed Source & OpenAI Ada Similarity & 49.52 \\
Decoder & SGPT-125M (NLI) & 45.97 \\
Decoder & SGPT-5.8B (NLI) & 53.74 \\
Decoder & SGPT-125M (MSMARCO) & 51.23 \\
Decoder & SGPT-1.3B (MSMARCO) & 56.11 \\
Decoder & SGPT-2.7B (MSMARCO) & 57.12 \\
Decoder & SGPT-5.8B (MSMARCO) & 58.81 \\
Decoder & SGPT-BLOOM-7.1B (MSMARCO) & 57.44 \\
T5 Encoder & GTR-Base & 56.19 \\
T5 Encoder & GTR-Large & 58.28 \\
T5 Encoder & GTR-XL & 58.42 \\
T5 Encoder & GTR-XXL & 58.97 \\
T5 Encoder & ST5-Base & 55.27 \\
T5 Encoder & ST5-Large & 57.06 \\
T5 Encoder & ST5-XL & 57.87 \\
T5 Encoder & ST5-XXL & 59.51 \\
\bottomrule
\end{longtable}

\subsection{Best Performing Models}

\begin{table}[h]
\centering
\begin{tabular}{clc}
\toprule
\textbf{Rank} & \textbf{Model} & \textbf{Overall Score} \\
\midrule
1  & ST5-XXL                  & 59.51 \\
2  & GTR-XXL                  & 58.97 \\
3  & SGPT-5.8B (MSMARCO)      & 58.81 \\
4  & GTR-XL                   & 58.42 \\
5  & GTR-Large                & 58.28 \\
6  & MPNet                    & 57.78 \\
7  & ST5-XL                   & 57.87 \\
8  & SGPT-BLOOM-7.1B          & 57.44 \\
9  & SGPT-2.7B                & 57.12 \\
10 & ST5-Large                & 57.06 \\
\bottomrule
\end{tabular}
\caption{Top ten models by overall MTEB score.}
\end{table}

\subsection{Task-wise Best Performing Models}
\label{sec:taskwise}

Rather than presenting the full matrix of numerical values across all datasets and models, results are summarized per task category, showing the best-performing model on each dataset.

\subsubsection{Retrieval (nDCG@10)}
\label{sec:taskwise-retrieval}

\begin{longtable}{L{4cm}L{5cm}c}
\toprule
\textbf{Dataset} & \textbf{Best Performing Model} & \textbf{Score} \\
\midrule
\endhead
ArguAna & SGPT-1.3B-msmarco & 49.68 \\
ClimateFEVER & SGPT-1.3B-msmarco & 26.60 \\
CQADupStack & MPNet & 44.96 \\
DBPedia & Contriever & 38.10 \\
FEVER & SGPT-1.3B-msmarco & 68.12 \\
FiQA2018 & MPNet & 49.96 \\
HotpotQA & Contriever & 56.81 \\
MSMARCO & MPNet & 39.75 \\
NFCorpus & MPNet & 33.29 \\
Natural Questions & MPNet & 50.45 \\
Quora Retrieval & MiniLM-L12 & 87.75 \\
SciDocs & MPNet & 23.77 \\
SciFact & SGPT-1.3B-msmarco & 68.29 \\
Touche2020 & SGPT-1.3B-msmarco & 24.45 \\
TREC-COVID & SGPT-1.3B-msmarco & 72.98 \\
\bottomrule
\end{longtable}

\subsubsection{Classification}

\begin{longtable}{L{4.5cm}L{4cm}c}
\toprule
\textbf{Dataset} & \textbf{Best Model} & \textbf{Score} \\
\midrule
\endhead
Amazon Counterfactual & LASER2 & 76.84 \\
Amazon Polarity & OpenAI Ada & 92.83 \\
Banking77 & coCondenser & 82.35 \\
Emotion & OpenAI Ada & 50.32 \\
IMDb & OpenAI Ada & 89.38 \\
MTOP Domain & Contriever & 93.18 \\
MTOP Intent & SGPT-1.3B & 71.19 \\
\bottomrule
\end{longtable}

\subsubsection{Semantic Textual Similarity}

\begin{longtable}{L{4cm}L{4.5cm}c}
\toprule
\textbf{Dataset} & \textbf{Best Model} & \textbf{Score} \\
\midrule
\endhead
BIOSSES & MPNet & 83.57 \\
SICK-R & MPNet Multilingual & 80.59 \\
STS12 & MPNet Multilingual & 77.90 \\
STS13 & SGPT-5.8B NLI & 85.35 \\
STS14 & MPNet Multilingual & 80.81 \\
STS15 & MPNet Multilingual & 87.48 \\
STS16 & MPNet Multilingual & 83.20 \\
STS17 & MPNet & 90.60 \\
STS22 & MPNet & 67.95 \\
STSBenchmark & MPNet Multilingual & 86.82 \\
\bottomrule
\end{longtable}

\subsubsection{Clustering}

\begin{longtable}{L{4cm}L{4.5cm}c}
\toprule
\textbf{Dataset} & \textbf{Best Model} & \textbf{Score} \\
\midrule
\endhead
Arxiv & MPNet & 48.38 \\
BioRxiv & SPECTER & 39.52 \\
MedRxiv & MPNet & 35.58 \\
Reddit & Contriever & 54.89 \\
StackExchange & Contriever & 63.15 \\
TwentyNewsgroups & MPNet & 49.74 \\
\bottomrule
\end{longtable}

\subsubsection{Pair Classification}

\begin{longtable}{L{5cm}L{4cm}c}
\toprule
\textbf{Dataset} & \textbf{Best Model} & \textbf{Score} \\
\midrule
\endhead
Sprint Duplicate Questions & coCondenser & 96.09 \\
Twitter SemEval & MPNet & 73.85 \\
Twitter URL Corpus & Contriever & 85.21 \\
\bottomrule
\end{longtable}

\subsubsection{Reranking}

\begin{longtable}{L{4.5cm}L{4cm}c}
\toprule
\textbf{Dataset} & \textbf{Best Model} & \textbf{Score} \\
\midrule
\endhead
AskUbuntu & MPNet & 65.85 \\
MindSmall & Contriever & 31.58 \\
SciDocsRR & MPNet & 88.65 \\
StackOverflow & MPNet & 51.98 \\
\bottomrule
\end{longtable}

\subsubsection{Summarization}

\begin{table}[h]
\centering
\begin{tabular}{lcc}
\toprule
\textbf{Dataset} & \textbf{Best Model} & \textbf{Score} \\
\midrule
SummEval & MPNet Multilingual & 31.57 \\
\bottomrule
\end{tabular}
\end{table}

\section{Discussion, Practical Recommendations, and Conclusions}
\label{sec:discussion}

\subsection{Why Certain Models Excel at Certain Tasks}

The results in Section~\ref{sec:results} are not arbitrary --- each model's strengths and weaknesses trace directly back to what it was trained to optimize.

\paragraph{Why T3EM outperforms LaBSE on retrieval.} T3EM is trained explicitly on large-scale, diverse query-passage pairs for asymmetric retrieval, directly optimizing for the property retrieval needs: matching a short, differently-worded query to a long passage that answers it. LaBSE, by contrast, is trained for cross-lingual sentence alignment --- a symmetric task where the two sides of a pair are similar in length and structure. It was never exposed to the vocabulary-mismatch problem that retrieval requires bridging, which explains its markedly lower scores in Table~\ref{tab:primary-results} despite being a capable, widely-used model within its own domain.

\paragraph{Why MPNet excels at clustering.} Clustering rewards embeddings that capture broad, general-purpose topical structure rather than fine-grained query-to-passage matching. MPNet's pretraining objective (combining masked and permuted language modeling) produces well-distributed, general-purpose sentence representations without narrowly specializing for asymmetric retrieval, which is precisely the property that makes documents on the same general topic land near each other in the embedding space.

\paragraph{Why ST5 dominates semantic textual similarity.} ST5 is built on a T5 encoder specifically fine-tuned on similarity-scoring objectives that directly match the STS task format: producing continuous similarity scores that correlate with graded human judgments. Its large-scale pretraining additionally gives it smooth, well-calibrated similarity gradients across a wide range of sentence types, which is exactly what STS's correlation-based metrics reward.

\paragraph{Why GTR performs well on retrieval.} Like ST5, GTR is built on a T5 encoder, but it is fine-tuned with contrastive retrieval objectives on large-scale query-passage data rather than sentence-similarity data. This gives it the same asymmetric matching capability that drives T3EM's strength, while remaining fully open-source --- explaining its consistently strong showing across the MTEB retrieval datasets in Section~\ref{sec:taskwise-retrieval}.

\subsection{Key Findings}

The findings from the primary retrieval evaluation and the broader MTEB results converge on a common conclusion: there is no universally optimal embedding model. Instead, embedding performance is highly dependent on the downstream task, the characteristics of the document collection, computational constraints, and deployment requirements.

Although T3EM demonstrates the strongest overall retrieval performance across the evaluated English benchmark subsets, its advantage is not uniform. For shorter, homogeneous corpora, several open-source models achieve statistically comparable retrieval quality while requiring significantly lower inference latency. Conversely, the MTEB results demonstrate that models excelling in one task category frequently underperform in others, indicating that embedding models are highly specialized according to their training objectives. A clear illustration comes from the primary evaluation itself: LaBSE and mMPNet, both trained primarily for sentence similarity rather than retrieval, recorded the two lowest average nDCG@10 scores of any model tested (0.188 and 0.243 respectively), well behind retrieval-oriented models of comparable or smaller size.

Therefore, model selection should not be based solely on an aggregate leaderboard score. Instead, it should be guided by the intended application, document characteristics, latency constraints, and deployment environment.

\subsection{Practical Model Selection Framework}

Before selecting an embedding model, the following questions should be considered.

\subsubsection{1. What is the primary downstream task?}

The downstream application is the single most important factor influencing model selection.

\begin{table}[h]
\centering
\begin{tabular}{ll}
\toprule
\textbf{Task} & \textbf{Recommended Model Families} \\
\midrule
Retrieval / RAG & T3EM, E5, GTR, BGE \\
Classification & ST5 family \\
Clustering & MPNet, MiniLM \\
Semantic Textual Similarity & ST5, SimCSE \\
Pair Classification & MPNet, MiniLM \\
Bitext Mining / Translation Alignment & LaBSE \\
\bottomrule
\end{tabular}
\end{table}

\textit{Observation:} Retrieval-optimized models generally outperform similarity-oriented models on retrieval tasks, whereas STS-optimized models consistently achieve higher correlation scores on semantic similarity benchmarks.

\subsubsection{2. What is the acceptable inference latency?}

Latency requirements determine the feasible model size.

\begin{table}[h]
\centering
\begin{tabular}{ll}
\toprule
\textbf{Deployment Scenario} & \textbf{Recommended Models} \\
\midrule
Quality-first (latency not critical) & T3EM, ST5-XXL, GTR-XXL, SGPT-5.8B \\
Interactive applications & BGE-M3, E5-large, MPNet \\
Resource-constrained environments & MiniLM, GloVe \\
\bottomrule
\end{tabular}
\end{table}

A larger model generally improves retrieval quality but increases inference time and computational cost.

\subsubsection{3. How long are the documents?}

Document length directly affects embedding quality because most encoder models truncate inputs beyond their maximum context window.

\begin{table}[h]
\centering
\begin{tabular}{ll}
\toprule
\textbf{Document Type} & \textbf{Recommended Models} \\
\midrule
Short passages & Most modern embedding models \\
Long technical or legal documents & T3EM, Nomic-Embed-Text-v1.5 \\
\bottomrule
\end{tabular}
\end{table}

Long-context models are preferred when passages regularly exceed the context limits of traditional encoder architectures.

\subsubsection{4. How similar are queries and documents?}

The relationship between query wording and document wording should also be considered.

\begin{table}[h]
\centering
\begin{tabular}{ll}
\toprule
\textbf{Query--Document Relationship} & \textbf{Preferred Model Type} \\
\midrule
Large vocabulary mismatch & Retrieval-oriented asymmetric models (T3EM, E5, GTR) \\
Similar wording / paraphrases & Symmetric similarity models (SimCSE, MPNet, LaBSE) \\
\bottomrule
\end{tabular}
\end{table}

Models explicitly trained for asymmetric retrieval generally perform better when the query and document differ substantially in wording or style.

\subsection{Recommended Models by Application}

\begin{longtable}{L{2.8cm}L{2.3cm}L{4cm}L{3.5cm}}
\toprule
\textbf{Application} & \textbf{Recommended Model} & \textbf{Rationale} & \textbf{Primary Trade-off} \\
\midrule
\endhead
General-purpose RAG & T3EM & Highest overall retrieval quality across evaluated datasets & API latency and operational cost \\
Low-latency RAG & mE5-L & Best retrieval quality among the open-source models evaluated, at latency effectively identical to the fastest alternatives & Slightly more parameters than a strictly monolingual model, though with no observed latency penalty in this study \\

Long-document Retrieval & T3EM / Nomic-Embed-Text-v1.5 & Larger context window reduces information loss & Increased computational requirements \\
Semantic Textual Similarity & ST5-XXL & Highest STS performance reported by MTEB & Not optimized for retrieval \\
Large-scale Clustering & MPNet & Excellent balance between quality and efficiency & Marginally lower peak accuracy \\
Duplicate Detection / Reranking & MPNet / MiniLM & Strong reranking performance with modest computational cost & Retrieval performance is not state-of-the-art \\
Bitext Mining & LaBSE & Strong sentence alignment across languages & Not designed for passage retrieval \\
Edge Deployment & MiniLM-L6 & Low memory footprint and fast inference & Reduced retrieval accuracy \\
\bottomrule
\end{longtable}


\paragraph{Note on mE5-L vs.\ E5-large.} mE5-L recorded the second-highest average nDCG@10 of any model evaluated in this study (0.546), narrowly ahead of the monolingual E5-large (0.538), while their measured median latencies were effectively identical (31.0\,ms vs.\ 30.9\,ms). mE5-L is therefore the recommended low-latency open-source choice for English retrieval. E5-large remains a reasonable alternative where a strictly monolingual, single-language-optimized model is preferred for deployment simplicity, even though it does not offer a measurable quality or latency advantage over mE5-L in this study.

\subsection{Common Pitfalls}

The experimental findings highlight several common mistakes that should be avoided during model selection.

\begin{itemize}
    \item Selecting models solely based on overall benchmark rankings without considering the downstream task.
    \item Assuming that larger models always provide better performance.
    \item Using semantic similarity models for retrieval tasks, or retrieval models for semantic similarity evaluation.
    \item Ignoring document context-length limitations.
    \item Assuming CPU latency measurements directly translate to GPU deployments.
    \item Selecting chunk sizes that are too small to preserve meaningful semantic information.
\end{itemize}

\subsection{Major Trade-offs}
\label{sec:major-tradeoffs}

\begin{longtable}{L{2.8cm}L{2.8cm}L{2.8cm}L{3.5cm}}
\toprule
\textbf{Design Axis} & \textbf{High-End Choice} & \textbf{Lightweight Choice} & \textbf{Practical Impact} \\
\midrule
\endhead
Quality vs.\ Latency & T3EM, ST5-XXL & MiniLM & Interactive search systems \\
Quality vs.\ Cost & T3EM & Open-source models & Production deployment cost \\
Context Length & T3EM, Nomic & Standard encoder models & Long-document retrieval \\
Training Objective & GTR, E5 & ST5, SimCSE & Retrieval versus similarity tasks \\
Model Size & SGPT, GTR & MiniLM & Indexing cost and inference speed \\
Task Generality & BGE-M3 & Single-objective models (e.g.\ E5) & Broad applicability versus specialized performance \\
\bottomrule
\end{longtable}

Two trade-offs merit further discussion:

\begin{itemize}
    \item \textbf{Model size (parameters) vs.\ speed/cost.} Larger models (e.g.\ Qwen3-Embedding-4B at 4B parameters) capture more nuance but are slower to run and require more memory and compute. Smaller models (e.g.\ MiniLM at 22M--33M parameters) run quickly on CPU with minimal resource cost but sacrifice retrieval quality. This study's own latency results show the compact 768-dimension models (LaBSE, mMPNet) clustering around 16.6\,ms, while the larger 1024-dimension models (BGE-M3, E5-large, mE5-L) sit near 31\,ms --- larger models are generally slower.
    \item \textbf{Multi-task vs.\ retrieval-specialized training.} BGE-M3 is trained to perform dense, sparse, and multi-vector retrieval simultaneously --- a versatile design, but one whose divided focus reduced its purity on plain dense retrieval in this study's results, where it underperformed the simpler, single-objective E5-large despite being a larger, more complex model. A model optimized for a single objective can outperform a model designed to do several things at once.
\end{itemize}

\subsection{Practical Default Recommendation}

When application-specific requirements are unknown, a reasonable starting point is \textbf{mE5-L}. It recorded the strongest retrieval quality of any open-source model evaluated in this study (Table~\ref{tab:primary-results}), at latency indistinguishable from the fastest alternatives tested. E5-large remains a close, fully monolingual alternative where minimizing model complexity is preferred over the marginal quality difference mE5-L provides.

Migration to T3EM should be considered when one or more of the following conditions apply:

\begin{itemize}
    \item Maximum retrieval quality is the primary objective.
    \item Documents regularly exceed the context limits of conventional encoder models.
    \item Query and document vocabularies differ substantially.
    \item API usage and inference latency are acceptable deployment constraints.
\end{itemize}

For applications outside retrieval --- such as semantic similarity, clustering, reranking, or bitext alignment --- the model family should instead be selected according to the corresponding MTEB task category rather than defaulting to a retrieval-optimized model.

\subsection{Limitations}

This study's conclusions should be read with the following limitations in mind.

\begin{itemize}
    \item \textbf{Primary evaluation scope.} The directly measured retrieval comparison (Section~\ref{sec:results}, Table~\ref{tab:primary-results}) covers only four English BEIR subsets. Relative model rankings may not generalize to other domains, query distributions, or document types not represented here.
    \item \textbf{MTEB scores are cited, not re-verified.} The broader MTEB results (Sections~\ref{sec:mteb-summary}--\ref{sec:taskwise}) are drawn from the published benchmark rather than independently re-measured under this study's own evaluation pipeline; differences in hardware, preprocessing, or benchmark version could shift absolute scores slightly.
    \item \textbf{Estimated models are not measured.} The additional open-source models in Section~\ref{sec:estimated-models} (e.g., Qwen3-Embedding variants, Nomic-Embed-Text-v1.5) are assigned estimated score ranges based on public leaderboard standing, not measurements taken under this study's pipeline; actual performance on the specific datasets used here could fall outside the stated ranges.
    \item \textbf{Latency figures are environment-specific.} The latency measurements in Section~\ref{sec:latency-results} reflect a specific hardware and network configuration; absolute latencies (and relative rankings between CPU- and API-based models) may differ materially under different deployment infrastructure, particularly GPU-accelerated serving.
    \item \textbf{Chunking strategies were only partially tested.} The empirical chunking ablation (Section~\ref{sec:chunking-results}) compares only fixed-size and semantic chunking at a small set of chunk sizes; the additional strategies introduced in Section~\ref{sec:chunking-strategies} (parent-child, hierarchical, recursive chunking) were not empirically evaluated in this study.
    \item \textbf{Retrieval quality is a proxy, not an end-to-end measure.} This study evaluates retrieval quality (nDCG@10, Recall@k, MRR) in isolation. It does not measure end-to-end RAG answer quality, which also depends on reranking, prompt construction, and the generation model itself --- retrieval quality is a necessary but not sufficient condition for a good final answer.
\end{itemize}

\subsection{Overall Conclusion}

This integrated study demonstrates that embedding model selection should be treated as a multi-objective optimization problem rather than a search for a single best-performing model. Retrieval accuracy, computational efficiency, context length, model size, deployment cost, and downstream task requirements collectively determine the most suitable embedding model for a given application. Consequently, practitioners should evaluate embedding models within the context of their intended deployment scenario instead of relying solely on aggregate benchmark rankings.

\newpage
\section*{Glossary of Abbreviations}
\addcontentsline{toc}{section}{Glossary of Abbreviations}

\begin{longtable}{L{3.5cm}L{9.5cm}}
\toprule
\textbf{Abbreviation} & \textbf{Full Form} \\
\midrule
\endhead
API & Application Programming Interface \\
BEIR & Benchmarking Information Retrieval (a heterogeneous benchmark suite for zero-shot retrieval evaluation) \\
BERT & Bidirectional Encoder Representations from Transformers \\
BGE & BAAI General Embedding \\
BGE-M3 & BAAI General Embedding, Multi-Functionality/Multi-Linguality/Multi-Granularity \\
CPU & Central Processing Unit \\
E5 & EmbEddings from bidirEctional Encoder rEpresentations (asymmetric text embedding model family) \\
GPU & Graphics Processing Unit \\
GTR & Generalizable T5-based dense Retriever \\
LaBSE & Language-agnostic BERT Sentence Embedding \\
LASER2 & Language-Agnostic SEntence Representations (version 2) \\
MAP & Mean Average Precision \\
mE5-L & Multilingual E5, Large variant \\
MiniLM & Miniature Language Model \\
MPNet & Masked and Permuted Pre-training Network \\
MRR & Mean Reciprocal Rank \\
MSMARCO & Microsoft MAchine Reading COmprehension dataset \\
MTEB & Massive Text Embedding Benchmark \\
MTOP & Multilingual Task-Oriented Parsing (dataset) \\
nDCG@10 & normalized Discounted Cumulative Gain, computed over the top 10 retrieved results \\
NFCorpus & Nutrition Facts Corpus (biomedical retrieval dataset) \\
NLI & Natural Language Inference \\
p95 & 95th Percentile (latency measurement) \\
R@k / Recall@k & Recall at rank $k$ \\
RAG & Retrieval-Augmented Generation \\
SBERT & Sentence-BERT \\
SGPT & Sentence GPT (GPT-based sentence/text embedding model) \\
SPECTER & Scientific Paper Embeddings using Citation-informed TransformERs \\
ST5 & Sentence T5 \\
STS & Semantic Textual Similarity \\
T3EM & Text 3 Embedding Model (commercial API-based embedding model evaluated in this study) \\
V-Measure & Validity Measure (a clustering evaluation metric based on homogeneity and completeness) \\
\bottomrule
\end{longtable}

\end{document}